\documentclass[runningheads]{llncs}

\usepackage[T1]{fontenc}

\usepackage{todonotes}
\usepackage{graphicx}
\usepackage{xcolor}
\usepackage{amsmath,mathtools,amssymb}
\usepackage{stmaryrd}
\usepackage[hidelinks]{hyperref}

\usepackage{listings}

\definecolor{primary}{RGB}{0, 51, 102}    
\definecolor{secondary}{RGB}{230, 240, 255} 
\definecolor{accent}{RGB}{100, 100, 100}  

\definecolor{keywordcolor}{rgb}{0.7, 0.1, 0.1}   
\definecolor{tacticcolor}{rgb}{0.0, 0.1, 0.6}    
\definecolor{commentcolor}{rgb}{0.4, 0.4, 0.4}   
\definecolor{symbolcolor}{rgb}{0.0, 0.1, 0.6}    
\definecolor{sortcolor}{rgb}{0.1, 0.5, 0.1}      
\definecolor{attributecolor}{rgb}{0.7, 0.1, 0.1} 

\usepackage{xspace}
\newcommand{\cslib}{CSLib\xspace}

\newcommand{\lto}{\xrightarrow}

\newcommand{\proptrue}{\text{true}}
\newcommand{\propfalse}{\text{false}}
\newcommand{\propand}[2]{#1 \wedge #2}
\newcommand{\propor}[2]{#1 \vee #2}
\newcommand{\propdiamond}[2]{\langle #1 \rangle #2}
\newcommand{\propbox}[2]{[#1]#2}

\begin{document}
\title{Hennessy--Milner Logic in \cslib, the Lean Computer Science Library}
%
%
\author{Fabrizio Montesi\inst{1}\orcidID{0000-0003-4666-901X} \and
Marco Peressotti\inst{1}\orcidID{0000-0002-0243-0480} \and
Alexandre Rademaker\inst{2}\orcidID{0000-0002-7583-0792}}
\authorrunning{F. Montesi et al.}
%
\institute{FORM, University of Southern Denmark, Odense, Denmark\\
\email{\{fmontesi,peressotti\}@imada.sdu.dk} \and
School for Applied Mathematics, Getulio Vargas Foundation, Rio de Janeiro, Brazil\\
\email{alexandre.rademaker@fgv.br}}
\maketitle              
\begin{abstract}
We present a library-level formalisation of Hennessy--Milner Logic (HML) -- a foundational logic for labelled transition systems (LTSs) -- for the Lean Computer Science Library (\cslib).
Our development includes the syntax, satisfaction relation, and denotational semantics of HML, as well as a complete metatheory including the Hennessy--Milner theorem -- bisimilarity coincides with theory equivalence for image-finite LTSs.
Our development emphasises generality and reusability: it is parametric over arbitrary LTSs, definitions integrate with \cslib's infrastructure (such as the formalisation of bisimilarity), and proofs leverage Lean's automation (notably the \texttt{grind} tactic).
All code is publicly available in \cslib and can be readily applied to systems that use its LTS API.

\keywords{Hennessy--Milner Logic \and Transition Systems \and Bisimulation}
\end{abstract}
\section{Introduction}
Labelled Transition Systems (LTSs) are the go-to framework for defining the operational behaviour of many computational systems, including process calculi, communication protocols, automata, and reactive systems~\cite{Sangiorgi_2011}.
Hennessy--Milner Logic (HML)~\cite{ref_hm85} is a fundamental modal logic for reasoning about such systems.
Its distinguishing feature is the \emph{Hennessy--Milner theorem}: over image-finite LTSs, two states satisfy exactly the same HML propositions if and only if they are strongly bisimilar.
This tight correspondence makes HML a very useful logical foundation in many contexts, from concurrency theory to model checking.

Despite its foundational status, mechanised treatments of HML at library-grade generality are scarce.
Existing formalisations tend to treat variations of HML or specific kinds of LTSs.
In this paper we report on a new formalisation of HML developed in Lean as part of the Lean Computer Science Library (\cslib) -- a rapidly-growing community project aimed at providing a universal basis for formalised computer science research and verified software development~\cite{cslib}.

Our main contributions are: 
%
%
(1)~universe-polymorphic definitions of the syntax, satisfaction relation, and denotational semantics of HML;
(2)~the complete fundamental metatheory of HML, including correctness of its denotational semantics and the Hennessy--Milner theorem (bisimilarity coincides with theory equivalence for image-finite LTSs);
(3)~integration into \cslib with documentation and appropriate APIs for downstream use.
For example, our development applies directly to automata theory and the Calculus of Communicating Systems (CCS)~\cite{Milner80} defined in \cslib{} -- those being based on the same general LTS API.

In the remainder, we assume some familiarity with theorem provers.

\noindent\textit{Code and reproducibility.} Our development is publicly available in the official \cslib repository.\footnote{\href{https://github.com/leanprover/cslib}{github.com/leanprover/cslib}, namespace \lstinline|Cslib.Logic.HML| at the time of writing.}
All code was checked with Lean version 4.28.0-rc1.

\section{Background: Labelled Transition Systems and Bisimilarity in \cslib}

\cslib offers a rich library for operational semantics. We recall the most important elements connected to our developments.

\begin{definition}[Labelled Transition System]
A \emph{labelled transition system} is a triple $(S, L, \to)$ where $S$ is the set of \emph{states}, $L$ is the set of {transition labels}, and $\to \mathop{\subseteq} S \times L \times S$ is the \emph{transition relation}.
We write $s \lto\mu s'$ for $(s, \mu, s') \in \to$.
\end{definition}
In \cslib, this is formalised as a structure parameterised over the types of states and labels, \lstinline|LTS State Label|, which contains the transition relation as a predicate \lstinline|Tr : State → Label → State → Prop|. (We write Lean code in \lstinline{fixed width}.)

\begin{definition}
The \emph{image} of a state $s$ for a transition label $\mu$ is the set of all states that $s$ can reach with a $\mu$-transition:
$\mathsf{image}(s,\mu) = \{s' \mid s \lto{\mu} s'\}$.
\end{definition}
The definition in \cslib is identical, thanks to the notation for sets offered by Mathlib~\cite{mathlib} (the Lean mathematical library, which \cslib is based on).
\begin{lstlisting}
def LTS.image (lts : LTS State Label) (s : State) (μ : Label) : Set State := { s' : State | lts.Tr s μ s' }
\end{lstlisting}

\begin{definition}[Image-Finiteness]
An LTS is \emph{image-finite} if for every state $s$ and label $\mu$, the set $\mathsf{image}(s,\mu)$ is finite.
\end{definition}
In \cslib, this is expressed as a typeclass for \lstinline|LTS| with the requirement \lstinline|∀ s μ, Finite (lts.image s μ)|.

\begin{definition}[Bisimulation and Bisimilarity]
A relation $R \subseteq S \times S$ is a \emph{bisimulation} if whenever $s_1 \mathbin R s_2$:
\begin{enumerate}
\item $s_1 \lto{\mu} s_1'$ for any $s_1'$ implies $s_2 \lto{\mu} s_2'$ for some $s_2'$ such that $s_1' \mathbin R s_2'$;
\item $s_2 \lto{\mu} s_2'$ for any $s_2'$ implies $s_1 \lto{\mu} s_1'$ for some $s_1'$ such that $s_1' \mathbin R s_2'$.
\end{enumerate}
Two states $s_1$ and $s_2$ are \emph{bisimilar}, written $s_1 \sim s_2$, if there is a bisimulation $R$ such that $s_1 \mathbin R s_2$.
\end{definition}
In \cslib, a relation \lstinline|r : State → State → Prop| is a bisimulation in an LTS \lstinline|lts| if it respects the predicate \lstinline|lts.IsBisimulation r|. This predicate and bisimilarity are direct formalisations of the definition above.

\section{HML Formalisation in Lean}

We now describe the core components of our formalisation of HML: syntax of propositions, satisfaction (semantics in terms of transitions), denotational semantics, and theory equivalence.

\subsection{Propositions}

Given a set of transition labels $L$, the language of HML propositions is given by the grammar:
\begin{align*}
\varphi \Coloneqq\; & \proptrue \mid \propfalse \mid \propand{\varphi}{\varphi} \mid \propor{\varphi}{\varphi} \mid \propdiamond{\mu}{\varphi} \mid \propbox{\mu}{\varphi}
\qquad\text{where }\mu \in L.
\end{align*}

Propositions state properties about a given state.
The modality $\propdiamond{\mu}{\varphi}$ (`diamond') asserts the existence of a transition with label $\mu$ from the given state to a state that satisfies $\varphi$, while $\propbox{\mu}{\varphi}$ (`box') requires that all transitions with label $\mu$ from the given state are to states that satisfy $\varphi$.

We define propositions in Lean as an inductive type parametric in the label type (\lstinline|Label|, corresponding to $L$ above).
\begin{lstlisting}
inductive Proposition (Label : Type u) : Type u where
  | true | false
  | and (φ₁ φ₂ : Proposition Label) | or (φ₁ φ₂ : Proposition Label)
  | diamond (μ : Label) (φ : Proposition Label)
  | box (μ : Label) (φ : Proposition Label)
\end{lstlisting}
The definition is universe-polymorphic (\lstinline|u|), and thus works with any label type.

There are two popular approaches to the language of propositions in HML. The original does not have $\propfalse$, but includes a negation connective~\cite{ref_hm85}.
We follow instead the other approach, from later developments, which has no negation but includes $\propfalse$~\cite{Aceto1999}. This makes the definition of satisfaction an elegant, straightforward recursion (in the next section).
As usual, we can recover negation as a function that transforms any proposition in its negated form.
\begin{lstlisting}
def Proposition.neg (φ : Proposition Label) : Proposition Label :=
  match φ with | .true => .false | and φ₁ φ₂ => or φ₁.neg φ₂.neg
               | diamond μ φ => box μ φ.neg | /- ... -/
\end{lstlisting}

\subsection{Satisfaction Relation}

The satisfaction relation $s \models \varphi$ (`state $s$ satisfies proposition $\varphi$') is defined inductively as follows. The $\propfalse$ proposition has no introduction rule, so no state satisfies it.
\begin{align*}
s &\models \proptrue && \text{(no rule for \propfalse)}\\
s &\models \propand{\varphi_1}{\varphi_2} && \text{if } s \models \varphi_1 \text{ and } s \models \varphi_2 \\
s &\models \propor{\varphi_1}{\varphi_2} && \text{if } s \models \varphi_1 \text{ or } s \models \varphi_2 \\
s &\models \propdiamond{\mu}{\varphi} && \text{if } \exists s'. s \xrightarrow{\mu} s' \wedge s' \models \varphi \\
s &\models \propbox{\mu}{\varphi} && \text{if } \forall s'. s \xrightarrow{\mu} s' \Rightarrow s' \models \varphi
\end{align*}

Again, we formalise this as an inductive.
\begin{lstlisting}
inductive Satisfies (lts : LTS State Label) :
    State → Proposition Label → Prop where
  | true : Satisfies lts s .true
  | and : Satisfies lts s φ₁ → Satisfies lts s φ₂ →
          Satisfies lts s (.and φ₁ φ₂)
  | or₁ : Satisfies lts s φ₁ → Satisfies lts s (.or φ₁ φ₂)
  | diamond : lts.Tr s μ s' → Satisfies lts s' φ →
              Satisfies lts s (.diamond μ φ)
  | /- ... -/
\end{lstlisting}

Note that the informal presentation assumes a fixed LTS, which translates to a parameter in our formalisation.
We write \lstinline|lts.Satisfies s φ| or $s \models \varphi$ interchangeably when the LTS is clear from the context.

With satisfaction defined, we can prove the correctness of negation.
\begin{lstlisting}
theorem neg_satisfies : ¬Satisfies lts s φ.neg ↔ Satisfies lts s φ
\end{lstlisting}

\subsection{Denotational Semantics}

HML comes with a compositional \emph{denotational semantics}, mapping each proposition to the set of states that satisfy it.
\begin{align*}
\llbracket \proptrue \rrbracket &= S  \quad \text{(the set of all states)}&
\llbracket \propfalse \rrbracket &= \emptyset \\
\llbracket \propand{\varphi_1}{\varphi_2} \rrbracket &= \llbracket \varphi_1 \rrbracket \cap \llbracket \varphi_2 \rrbracket &
\llbracket \propor{\varphi_1}{\varphi_2} \rrbracket &= \llbracket \varphi_1 \rrbracket \cup \llbracket \varphi_2 \rrbracket \\
\llbracket \propdiamond{\mu}{\varphi} \rrbracket &= \{ s \mid \exists s'.\, s \xrightarrow{\mu} s' \land s' \in \llbracket \varphi \rrbracket \} &
\llbracket \propbox{\mu}{\varphi} \rrbracket &= \{ s \mid \forall s'.\, s \xrightarrow{\mu} s' \Rightarrow s' \in \llbracket \varphi \rrbracket \}
\end{align*}

This is straightforward to translate to Lean, as follows.
\begin{lstlisting}
def denotation (φ : Proposition Label) (lts : LTS State Label) :
    Set State := match φ with
  | .true => Set.univ | .false => ∅
  | .and φ₁ φ₂ => φ₁.denotation lts ∩ φ₂.denotation lts
  | .or φ₁ φ₂ => φ₁.denotation lts ∪ φ₂.denotation lts
  | .diamond μ φ => { s | ∃ s', lts.Tr s μ s' ∧ s' ∈ φ.denotation lts }
  | .box μ φ => { s | ∀ s', lts.Tr s μ s' → s' ∈ φ.denotation lts }
\end{lstlisting}

\subsection{Theories and Theory Equivalence}

Satisfaction yields the concepts of theory, theory equivalence (sometimes called `logical equivalence' in previous work), and distinguishing propositions.

\begin{definition}[Theory and Theory Equivalence]
The \emph{theory} of a state $s$ is the set of all propositions it satisfies:
$
\mathsf{theory}(s) = \{ \varphi \mid s \models \varphi \}
$.
Two states are \emph{theory equivalent}, written $\mathsf{TheoryEq}(s_1, s_2)$, if $\mathsf{theory}(s_1) = \mathsf{theory}(s_2)$.
\end{definition}

We translate this as:
\begin{lstlisting}
def theory (lts : LTS State Label) (s : State) :
    Set (Proposition Label) := { φ | lts.Satisfies s φ }
def TheoryEq (lts : LTS State Label) (s₁ s₂ : State) : Prop :=
  theory lts s₁ = theory lts s₂
\end{lstlisting}

A useful lemma for our development is that, for any two states that are not theory equivalent, there always exists a distinguishing proposition.
\begin{lstlisting}
lemma not_theoryEq_satisfies (h : ¬ TheoryEq lts s1 s2) :
    ∃ φ, (Satisfies lts s1 φ ∧ ¬Satisfies lts s2 φ)
\end{lstlisting}
Negation is useful in obtaining an elegant statement here -- without it, it would have to be \lstinline|∃ φ, (Satisfies lts s1 φ ∧ ¬Satisfies lts s2 φ) ∨ (¬Satisfies lts s1 φ ∧ Satisfies lts s2 φ)|, complicating proofs later.

\section{Metatheory}

We move to our key results, which formalise the metatheory of HML.

\subsection{Correctness of the Denotational Semantics}

The inductive satisfaction relation and the denotational semantics coincide.

\begin{theorem}[Semantic Equivalence]\label{thm:sem-equiv}
$s \models \varphi$ if and only if $s \in \llbracket \varphi \rrbracket$.
\end{theorem}
The Lean proof follows the same principle found in the literature -- we proceed by induction on the structure of the proposition.
(The curly brackets denote an implicit parameter, which Lean tries to find in the context during elaboration.)
\begin{lstlisting}
theorem satisfies_mem_denotation {lts : LTS State Label} :
    Satisfies lts s φ ↔ s ∈ φ.denotation lts := by
  induction φ generalizing s <;> grind
\end{lstlisting}
The \lstinline|grind| tactic can deal with all cases automatically because our code, Lean's library, and Mathlib provide sufficient infrastructure (annotations) for \lstinline|grind| to unfold our definitions and deal with all the necessary operations on sets.



\subsection{Bisimulation Invariance}

A key property of HML is that bisimilar states cannot be logically distinguished.

\begin{theorem}[Bisimulation Invariance]\label{thm:bisim-inv}
If $R$ is a bisimulation and $s_1 \mathbin R s_2$, then $s_1 \models \varphi$ implies $s_2 \models \varphi$.
\end{theorem}

The theorem statement in Lean reads as follows.
\begin{lstlisting}
lemma bisimulation_satisfies {hrb : lts.IsBisimulation r}
    (hr : r s1 s2) (φ : Proposition Label) (hs : Satisfies lts s1 φ) :
    Satisfies lts s2 φ
\end{lstlisting}
The proof is by induction on \lstinline|φ|. All cases are solved by \lstinline|grind|, requiring only a minor hint to follow the bisimulation in the diamond case.

With this theorem and the fact that bisimilarity is symmetric, we can prove that bisimulation implies theory equivalence.
\begin{corollary}[Bisimulation implies Theory Equivalence]\label{cor:bisim-theory}
If $R$ is a bisimulation and $s_1 \mathbin R s_2$, then $\mathsf{TheoryEq}(s_1, s_2)$.
\end{corollary}
\begin{lstlisting}
lemma bisimulation_TheoryEq {lts : LTS State Label}
    {hrb : lts.IsBisimulation r} (hr : r s1 s2) : TheoryEq lts s1 s2
\end{lstlisting}

\subsection{Characterisation of Bisimilarity (Hennessy--Milner Theorem)}

The converse of Corollary~\ref{cor:bisim-theory} -- that theory equivalence implies bisimilarity -- holds only for image-finite LTS.
This is because, with the connectives of HML, one can completely specify only a finite number of behaviours in the image of a state.
The proof proceeds by showing that theory equivalence is itself a bisimulation.

\begin{lemma}\label{lem:theoryeq-bisim}
In any image-finite LTS, $\mathsf{TheoryEq}$ is a bisimulation.
\end{lemma}
Using \cslib's API, we formulate the lemma as follows.
\begin{lstlisting}
lemma theoryEq_isBisimulation (lts : LTS State Label)
    [image_finite : ∀ s μ, Finite (lts.image s μ)] :
    lts.IsBisimulation (TheoryEq lts)
\end{lstlisting}

Proving this required more creativity than for the other results but we could still rely on a standard argument.
Assume that states $s_1$ and $s_2$ are theory equivalent, and consider for example the case where $s_1 \lto\mu s_1'$ for some $\mu$ and $s_1'$.
We proceed by showing that $s_2$ not being able to match this transition leads to a contradiction.
That is, we (absurdly) assume that every $\mu$-derivative of $s_2$ is distinguishable from $s_1'$.
By the fact that the LTS is image-finite, we can (noncomputably) collect all the states reached from $s_2$ via a $\mu$-transition, obtain a distinguishing proposition $\phi_i$ for each, and compose those into a proposition $\propdiamond{\mu}{\bigwedge_{i=1}^{n} \phi_i}$ (where $n$ is the cardinality of the image of $s_2$).
We then prove that this proposition distinguishes $s_1$ from $s_2$, hence reaching the contradiction. 

The Hennessy--Milner theorem follows immediately.
\begin{theorem}[Hennessy--Milner]\label{thm:hm}
For any image-finite LTS, $\mathsf{TheoryEq} \;=\; {\sim}$.
\end{theorem}
The Lean translation is immediate.
\begin{lstlisting}
theorem theoryEq_eq_bisimilarity (lts : LTS State Label)
    [image_finite : ∀ s μ, Finite (lts.image s μ)] :
    TheoryEq lts = Bisimilarity lts
\end{lstlisting}

All our definitions and results readily apply to other parts of \cslib and downstream projects that use the \lstinline|LTS| API.
For instance, we can prove that certain automata and concurrent processes satisfy HML propositions, and these systems inherit our metatheoretical results.
The next snippet exemplifies that theory equivalence coincides with bisimilarity in \cslib's CCS (which follows~\cite{Sangiorgi_2011}).

\begin{lstlisting}
example [∀ p μ, Finite ((CCS.lts (defs := d)).image p μ)] :
  TheoryEq (CCS.lts (defs := d)) = Bisimilarity (CCS.lts (defs := d)) :=
  theoryEq_eq_bisimilarity _
\end{lstlisting}
The example works for any set of CCS recursive procedure definitions \lstinline|d|, as long as it respects the image-finiteness condition (e.g., using guarded recursion~\cite{Sangiorgi_2011}).

\section{Related Work and Conclusion}

Variations and extensions of HML have been formalised in different theorem provers~\cite{M01,C03,TW07,HRF15}. Differently from here, these efforts do not focus on proving the general metatheoretical results of HML, like the Hennessy--Milner theorem.

The work nearest to ours formalises in Isabelle an extension of HML to deal with a new kind of transition systems,  called nominal transition systems, and includes a proof of the Hennessy--Milner theorem for this framework \cite{PBEGW21}.
By contrast, our development is faithful to the original and is much simpler, yet suffices to prove the metatheory of HML in its full generality (for any LTS).
Furthermore, our work integrates with the APIs found in \cslib, allowing for the direct application of our results to all projects that use the same APIs (e.g., automata theory and CCS as given in \cslib, as well as downstream projects).

The complications in \cite{PBEGW21} are motivated by the aim of reasoning on process calculi like the $\pi$-calculus and its extensions, where the derivative process in a transition can depend on variable names introduced by the transition label (bound variables).
Reasoning effectively about such processes requires extending HML to allow propositions to depend on the bound names of transition labels written in the diamond and box modalities.
We leave this and extensions aimed at other features (like other types of dependencies) to future work.
An important design question for \cslib is how a general framework for all extensions of HML targetting different calculi can be offered.
Interesting future work includes also extensions to weak equivalences and fixpoint operators.

Our development is the first modal logic formalisation in \cslib and serves as its current reference implementation of Hennessy--Milner Logic, on which future extensions and applications within the library can reliably build.

\begin{credits}
\subsubsection{\ackname}
We thank Christopher Henson for comments on our code.
Alexandre Rademaker gratefully acknowledges support from Schmidt Sciences. 
Co-funded by the European Union (ERC, CHORDS, 101124225). Views and opinions expressed are however those of the authors only and do not necessarily reflect those of the European Union or the European Research Council. Neither the European Union nor the granting authority can be held responsible for them.
\end{credits}
%
%
\bibliographystyle{splncs04}
\bibliography{biblio}
\end{document}